\input{aipcheck}

\documentclass[
    ,final            
  ]
  {aipproc}

\layoutstyle{6x9}

\begin{document}

\title{Investigation of subtreshold resonances
with the Trojan Horse Method}

\author{G.Baur}{ address={Institut f\"{u}r Kernphysik, Forschungszentrum J\"{u}lich, 
    D-52425 J\"{u}lich, Germany}
}

\author{S.Typel}{
  address={Gesellschaft f\"{u}r Schwerionenforschung mbH (GSI),
           Planckstra\ss{}e 1, D-64291 Darmstadt, Germany}
}

\begin{abstract}
It is pointed out that the Trojan-Horse method 
is a suitable tool to investigate subthreshold resonances.
\end{abstract}

\maketitle




\section{Transfer Reactions and Trojan Horse method}

A similarity between cross sections for two-body and closely
related three-body reactions under certain kinematical conditions
\cite{Fuc71}
led to the introduction of the Trojan-Horse method 
\cite{Bau86,Typ00,tyba02}.
In this indirect approach a two-body reaction
\begin{equation} \label{APreac}
 A + x \to C + c
\end{equation}
that is relevant to nuclear astrophysics is replaced by a reaction
\begin{equation} \label{THreac}
 A + a \to C + c + b
\end{equation}
with three particles in the final state. 
One assumes that the Trojan horse
$a$ is composed predominantly of clusters $x$ and $b$, i.e.\  $a=(x+b)$. 
This reaction can be considered as a special case of a transfer 
reaction to the continuum. It is studied experimentally under quasi-free
scattering conditions, i.e.\ when the momentum transfer to the
spectator $b$ is small. The method was primarily applied to the
extraction of the low-energy cross section of reaction
(\ref{APreac}) that is relevant for astrophysics. However, the method
can also be applied to the study of single-particle states in exotic
nuclei around the particle threshold.
The basic assumptions of the Trojan Horse Method are discussed
in detail in \cite{tyba02}, see also \cite{mukha}. 

It is the purpose of this contribution to study the transition from the 
bound ($E_{Ax}<0$) to the unbound ($E_{Ax}>0$) region. 
We study the case where there is an open channel
$c+C \neq A+x$ at the $E_{Ax}=0$ threshold. We show that there
is a continuous transition. If there is a subthreshold resonance 
in the $B=A+x$-system it can be experimentally studied in the 
$A+(b+x) \rightarrow C+c +b$ reaction. 

\section{Continuous Transition from Bound to Unbound State
Stripping}

 For the  case where only the elastic channel 
$A+x$ is open we refer to Ch. 4.2.1 of \cite{msuproc}.
Now we study the case where the reaction $x+A \rightarrow C+c$ has
a positive $Q$ value, the relative energy $E$ between 
A and x can be negative
as well as positive in the three-body reaction $ A+a (=b+x)
\rightarrow C+c+b$.
As an example we quote the recently studied Trojan horse reaction
$d + ^{6}Li \rightarrow \alpha + ^3He + n$ \cite{auro03} 
applied to the ${}^{6}$Li$(p,\alpha)^{3}$He
two-body reaction (the neutron being the spectator).
In this case there are two charged
particles in the initial state (${}^{6}$Li+$p$)
and the $\alpha + ^3$He-channel is open at the 
$E=E_{^6Li+p}=0$-threshold.
  
The general question which we want to answer
here is how the two regions $E>0$ and
$E<0$ are related to each other. 
In fig.1 
\begin{figure}
\includegraphics[height=.3\textheight]{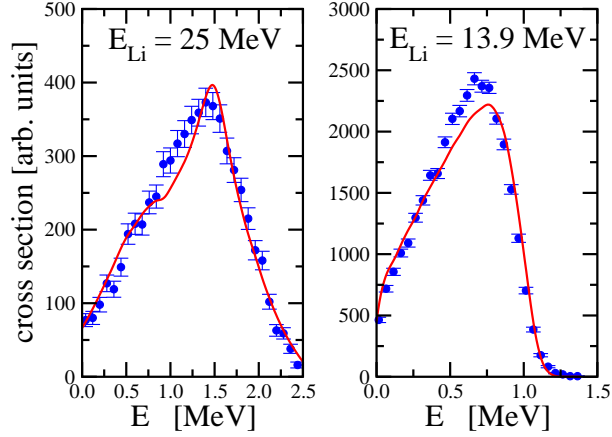}
\caption{\label{fig:1} The double differential 
cross section of the Trojan horse reaction
$d+^6Li \rightarrow \alpha + ^3He + n$
is finite at the $p+ ^6Li$ relative energy
$E=0$. How does it continue to
energies $E<0$?
}
\end{figure}
(fig.7 of \cite{auro03}) the coincidence yield 
is plotted as a function of the ${}^{6}$Li-$p$ relative energy $E$.
It is nonzero at zero relative energy. How does 
the theory \cite{tyba02} (and the experiment)
continue to negative relative energies?
With this method, subtreshold resonances can be 
investigated rather directly.

The cross section is a quantity which only
exists for $E>0$. However, a quantity like the S factor
(or related to it)
can be continued to energies below the threshold.
An instructive example is the modified shape function $\tilde{S}$ in Ch.\ 6
of \cite{tyba04}. In analogy to the astrophysical S factor,
where the Coulomb barrier is taken out,  the
angular momentum barrier is taken out
in $\tilde{S}$. As can be seen from table 3 or 4
of \cite{tyba04} $\tilde{S}$ is well defined 
for $x^2<0$, with the characteristic pole at $x^2=-1$,
corresponding to the binding energy of the $(A+x)=B$-system.

The inclusive breakup theory of IAV \cite{iav, mun} is extended to
negative energies $E<0$ in \cite{msuproc}.
We refer the reader to this reference for this approach. 

An alternative approach is based directly on the formulation of 
\cite{tyba02}. The Trojan horse cross section is given in eq.(61) there.
The S-matrixelement $S_{Ax Cc}$ can be found from the 
asymptotics of the radial wave functions, see eq. (23) there.
For negative energies $E<0$ the channel is closed and the outgoing
wave function is replaced by an exponentially decaying one.
The quantity $S_{Ax Cc}$ then is no longer an S-matrixelement,
but it plays the role of a normalization constant, 
see also \cite{wigner,lanethom}.

The Trojan horse amplitude involves the combination 
$f \propto S_{lc} \cdot J^+_l$
(see eq. (61) of \cite{tyba02}.
The threshold behaviour of $J^+$ in the case of neutrons is given
by A.30, for charged particles by eq. 59 of \cite{tyba02}. The energy 
behaviour of the inelastic $cC\neq Ax$ S-matrixelement is found from
the behaviour of the elastic S-matrixelement $S_{ll}$ and 
unitarity: $|S_{lc}|^2= 1-|S_{ll}|^2$ where $l \neq c$.
We have $S_{ll}=\exp 2i \delta _l \sim 1-2i \delta_l^R -2 \delta_l^I$,
where the imaginary part of the phase shift goes like $ \delta_l^I
\sim (kR)^{2l+1}$ for neutral particles. This
leads to $|S_{lc}|^2 \sim (kR)^{2l+1}$.
The product $S_{lc} \cdot  J_l^+$ is given by
$|S_{lc} J_l^+|^2 \sim (kR)^3$  and the cross section 
$\sigma \propto \frac{1}{k^3}|S_{lc} J_l^+|^2$ tends to a 
finite constant independent of k.
A similar analysis can be done for charged particles.

We now study the smooth change from $E>0$ to $E<0$ in
a two-channel model with a surface coupling. For the case
of $l=0$ it was studied in Ch. 4.2.3 of \cite{msuproc}.
Two radial functions $f_1$ and $f_2$ (we assume the 
same l-value) are coupled by a potential of the type
$u_{ij}(r) f_j (r)$. We take $u_{ij}=Q \delta(r-R)$. 
We are especially interested in the case where there is a resonance
just below or above the threshold of channel 1$\equiv (A+x)$,
channel 2 ($\equiv C+c)$ is always open.

In the channel 1 the wave function is $\zeta =N u_l^+ (iqr)$, for $E<0$
and $\zeta = S u_l^+(kr)$, for $E>0$. 
The in- and outgoing wave functions are given by
$u_l^\pm =e^{\mp i  \sigma_l}(G_l \pm iF_l)\rightarrow
\exp(\pm i (kr -\eta ln(2kr) -\frac{l \pi}{2}))$. For $E<0$
they correspond to exponentially decreasing and increasing 
wave functions.
The logarithmic derivative
in this channel is independent of $S$ or $N$.
For $r \geq R$ the radial wave functions are given in terms of the
S-matrix-elements by
\begin{equation}
f_1 =\sqrt{k_2/k_1}\frac{1}{2i}  S_{12} u_l^+(k_1r)
\end{equation}
and 
\begin{equation}
f_2 =\frac{1}{2i}(S_{22} u_l^+(k_2r)- u_l^- (k_2 r))
\end{equation} 
The logarithmic matching conditions
lead to a 'Sprungbedingung', which determines the S-matrixelements.
Denoting the interior logarithmic derivatives ($L\equiv \frac{r f{'}}{f}$)
by $L_1$ and $L_2$
we have  
\begin{equation}
\frac{ R f_1^{'}(R_>)}{f_1(R_>)}\equiv q_l^+(\kappa_1)=L_1 +Q R \frac{f_2(R)}{f_1(R)}
\end{equation}
and 
\begin{equation}
\frac{R f_2^{'}(R_>)}{f_2(R_>)}=L_2 +Q R \frac{f_1(R)}{f_2(R)}
\end{equation}
where $\kappa_1\equiv k_1 R$. Eq. (5) can be solved for $f_1$. The LHS is a 'kinematic '
quantity, it is given by $q_l^+\equiv \frac{\kappa
u_l^{+ '}}{u_l^+}$
and we have 
\begin{equation}
f_1(R)=\frac{ Q R f_2}{q_l^+(\kappa_1) - L_1} 
\end{equation}
For the case of neutral particles (neutrons)
we have the following behaviour close to threshold : 
$\rm{Re} q_l^+ =-l + O(k^2)$ and
$\rm{Im} q_l^+ \sim (\kappa_1)^{2l+1}/((2l-1)!!)^2 \equiv s_l
\equiv \kappa_1 v_l$. The penetration factor
$s_l$ is a small number. Charged particles can be treated in
an analogous way.
Inserting into eq.(9) we obtain an equation for the unknown S-matrixelement $S_{22}$:
\begin{equation}
\kappa_2 \frac{S_{22}\frac{d u_l^+}{d \kappa_2} -
\frac{d u_l^-}{d \kappa _2}}{S_{22}u_l^+-u_l^-}=
L_2 + \frac{(QR)^2}{q_l^+(\kappa_1) -L_1} \equiv \tilde{L_2} 
\end{equation}
where $\kappa _2 \equiv k_2 R$.  This equation can be solved for $S_{22}$,
we write $S_{22}=\exp(2 i \tau_l) S_{22}^{res}$,
where $\tau _l $ is the hard sphere phase shift.
We find 
\begin{equation}
S_{22}^{res}=\frac{\tilde{L_2} - q_l^-(\kappa_2)}
{\tilde{L_2} -q_l^+(\kappa_2)}
\end{equation}
 If channel 1 is closed,   $q_l^+ (\kappa_1)$ is real, thus  
$\tilde{L_2}$ is real and 
one obtains the form of a single particle resonance
with appropriate parameters. The S-matrix is $1 \times 1$  and $S_{22}$ is
unitary. If channel 1 is open: $\tilde{L_2}$ is complex and we can
bring $S_{22}^{res}$ in 
a two-channel Breit- Wigner resonance form. $S_{12}$
is then obtained (for a closed or an open channel 1) from eqs. 3 and 7. 

Although our model is quite special,
the resulting form has a general validity, with 
('effective') resonance parameters
$E_R$ (position)  and partial widths $\Gamma_1, \Gamma_2$. The
total width is $\Gamma= \Gamma_1 + \Gamma_2$.
It is a merit of our  model that it shows directly how
$S_{12}$ is extended to closed channels. It will be interesting to elucidate
the relation of the present model to the more general
approach of Ref. \cite{mutri}. 

The standard Breit-Wigner result is (see e.g.\cite{rr}) 
\begin{equation}
S_{ij}=\exp{i(\xi_i + \xi_j)}(\delta_{ij}-\frac{i \sqrt{\Gamma_i \Gamma_j}}
{E-E_R +i \Gamma/2})
\end{equation}
%

We now show  how to obtain these
forms from the coupled channel model. The condition for a resonance
at position $E_R$ is
$\rm{Re} \tilde{L_2}-\rm{Re} q_l^+(E_R)=0$: This defines $E_R$. 
In the resonance energy
region we can write $\rm{Re} \tilde{L_2}-\rm{Re} q_l^+=-c_3 (E-E_R)$
where $c_3$ is some constant.
The partial widths $\Gamma_1$ and $\Gamma_2$ are 
found from $\frac{1}{2}( \Gamma_1 + \Gamma_2) \equiv 
\frac{-1}{2c_3} (\rm{Im} \tilde{L_2} - \rm{Im} q_l^+ (\kappa_{2,R}))$ and $\frac{1}{2}
 (\Gamma_1 -\Gamma_2)
\equiv \frac{-1}{2c_3}(\rm{Im} \tilde{L_2} - \rm{Im} q_l^- (\kappa_{2,R}))$.
We assume that we have a single particle resonance in the uncoupled channel 2,
whereas there is no resonance structure in channel 1 (i.e. $\rm{Re} q_l^+-L_1$
is different from zero).
The 'bare' resonance condition is  $ L_2 (=\rm{real}) = \rm{Re} q_l^+(\kappa_2)$.
The coupling induces a shift of the resonance energy.

We have $\rm{Im} q_l^+=- \rm{Im} q_l^-(\kappa)= s_l(\kappa)$.
Since $s_l(\kappa_1)<<1$ we have $\rm{Im} \tilde{L_2} = 
\frac{-(QR)^2 s_l}{(l+L_1)^2}$.
We obtain 
\begin{equation}
\Gamma_1= \frac{2s_l(\kappa_1)(QR)^2}{c_3 (l+L_1)^2} 
\end{equation}
and 
\begin{equation}
\Gamma_2=\frac{2 s_l(\kappa_2)}{c_3}
\end{equation}
Thus the Breit-Wigner form of $S_{22}$ is recovered.
We see that $\Gamma_1$ is strongly energy dependent:
it contains the threshold penetration
factor $s_l(\kappa_1)$, whereas $\Gamma_2$ 
(and thus also the total width $\Gamma$)
do not. For $E \leq 0$ 
the S-matrix 
consists only of $S_{22}$, which is unitary. Still, $S_{12}$
which now plays the role of a normalization factor is nonzero.

By direct calculation we can bring $S_{12}$ in the Breit-Wigner form:
We write $ f_2(\kappa_2)=S_{22}u_l^+ - u_l^- =u_l^- (S_{22}^{res}-1)$.
 From eqs. (3,4) and (7) we find
\begin{equation}
S_{12}= \sqrt{k_1/k_2} \frac{-QR u_l^-(\kappa_2)}{u_l^+(\kappa_1)(l+L_1)}
\frac{-i \Gamma_2}{E-E_R+\frac{i}{2} (\Gamma_1 + \Gamma_2)}
=e^{i(\tau_1 + \tau_2)} \frac{-i \sqrt{\Gamma_1 \Gamma_2}}{E-E_R +\frac{i}{2}
(\Gamma_1 + \Gamma_2)}
\end{equation}
We used  eqs. (11) and (12), $u_l^\pm =\frac{1}{\sqrt v_l} e^{\mp i \tau_l}$
and $s_l \equiv \kappa v_l$to obtain this result.

This result is valid for $E>0$.
For $E<0$ there are some 
simple modifications. The wave number becomes imaginary, $k=iq$.
The quantity $q_l^\pm$ is real, $u_l^+$ can again be written 
as $1/\sqrt{v_l} \exp{\frac{-i\pi l}{2}}$(for neutral particles).
Thus we have $\tau_1 = \frac{i\pi l}{2} $ for $E<0$. 
Formally, the penetration factor $s_l$ turns imaginary
($s_l =i (-1)^{2l}(qR)^{2l+1}/((2l-1)!!)^2)$, 
and so does the width
$\Gamma_1$, see eq. (11).

 For $E_R>0$ we have a resonance, for $E_R<0$ a subthreshold resonance,
the formulae are valid for both cases.
The opening of channel 1 at threshold will induce cusp effects
in the elastic $C+c \rightarrow C+c$ scattering, see e.g. \cite{newton} .


\subsection{Large $p_{1/2}$-scattering length in the $^{11}$Be
system due to a neutron halo state}

The electromagnetic dipole strength in $^{11}$Be was deduced 
\cite{tybaprl} from high-energy
${}^{11}$Be Coulomb dissociation messurements at GSI \cite{palit}.
Using a cutoff radius of $R=2.78$~fm and 
an inverse bound-state decay length of
$q=0.1486$~fm${}^{-1}$ as input parameters we extract
an ANC of $C_{0}=0.724(8)$~fm${}^{-1/2}$ 
from the fit to the experimental data. The ANC
can be converted to a spectroscopic factor of $C^{2}S=0.704(15)$
that is consistent with results from other methods.
In the lowest order of 
the effective-range expansion the phase shift 
$\delta_{l}^{j}$
in the partial wave with orbital angular momentum $l$ and
total angular momentum $j$
is written as $\tan \delta_{l}^{j}= -a_l k^{2l+1}= -(x c_{l}^{j}\gamma)^{2l+1}$, 
where $\gamma=qR=0.4132<1$ is the halo expansion parameter and $x=k/q
=\sqrt{E/S_{n}}$. The neutron separation energy is $S_{n}$. 
The parameter  $c_{l}^{j}$ corresponds to the scattering
length $a_{l}^{j} = (c_{l}^{j}R)^{2l+1}$. We obtain
$c^{3/2}_{1}=-0.41(86,-20)$ and $c^{1/2}_{1}=2.77(13,-14)$. The latter 
is unnaturally large because of the existence of a bound $\frac{1}{2}^{-}$
state close to the neutron breakup threshold in ${}^{11}$Be.

 The connection of the scattering length $a_l$ and the bound state
parameter q for $l>0$ is given by
$a_l=\frac{2 (2l-1) R^{2l-1}}{q^2(2l+1)!!(2l-1)!!}$,
where a square well potential model with a range R was assumed.
This is a generalization of the
well-known relation $a_0=1/q$ for $l=0$.
The $p_{1/2}$ channel in $^{11}$Be is an example for the 
influence of a halo state on the continuum. 
The binding energy
of this state is given by 184 keV, which corresponds to $q=0.094$~fm. 
With $R=2.78$~fm one has $\gamma^2=0.068$.
For $l=1$ one has $a_1=\frac{2R^3}{3\gamma^2}=210$~ fm$^3$ 
which translates into $c_1= (a_1/R^3)^{1/3}=2.14$. This compares
favourably with the fit value given in table 1 of \cite{tybaprl}: 
$2.77(13,-14)$. The corresponding scattering length is given by 
$a_1^{1/2}=457(67,-66) fm^3$.
For a further discussion we refer to \cite{tybaprl}. The large 
$p_{1/2}$-scattering length would also manifest itself in 
the $^{10}Be(d,p\gamma)^{11}Be$ 'radiative transfer reaction'.


\section{Conclusion and Outlook}

The treatment of the continuum is a general problem,
which becomes more and more urgent when the dripline is approached.
We studied 
the transition from bound to unbound states
as a typical example. In the present analysis this transition
is continuous, it is expected that this also shows up in the experimental data.
A minireview of the applications of the Trojan horse method 
to astrophysical reactions can be found in  \cite {mumbai}
An extension to 
stripping into the continuum would be of interest for this 
and other kinds of reactions. The trojan horse reaction
$^{18}F(d,n \alpha)^{15}O$ would be of great interest for the 
$^{18}F(p,\alpha)^{15}O$ reaction relevant 
for nova nucleosynthesis \cite{nucnews,hribf}. This would also be of interest for 
SPIRAL2, for example. Another most interesting example would be the 
radiative $\alpha$-transfer reaction $^{12}C(^6Li,d \gamma)^{16}O$,
where one could also look directly for the subtreshold $1^-$ and $2^+$
states which are crucial for the S-factor of 
the astrophysically important $\alpha$-capture reaction $^{12}C(\alpha,\gamma)^{16}O$.




\bibliographystyle{aipproc}   


%

\end{document}